\documentclass[12pt]{iopart}

\usepackage{amssymb}
\usepackage{psfrag}
\usepackage{amstext}
\usepackage{epsfig}
\usepackage{longtable}
\usepackage{multirow}
\usepackage{tabularx}
\usepackage[verbose]{layout}
\usepackage{afterpage}
\usepackage{braket}

\usepackage{pstricks,pst-grad}

\begin{document}

\title{Broadband Optical Delay with Large Dynamic Range Using Atomic Dispersion}

\author{M. R. Vanner, R. J. McLean, P. Hannaford and A. M. Akulshin}

\address{ARC Centre of Excellence for Quantum-Atom Optics, Centre for Atom Optics and Ultrafast Spectroscopy, Swinburne
University of Technology, Melbourne, Australia 3122}


\begin{abstract}
We report on a tunable all-optical delay line for pulses with
optical frequency within the Rb $D_2$ absorption line. Using
frequency tuning between absorption components from different
isotopes, pulses of 10~ns duration are delayed in a 10~cm hot vapour
cell by up to 40~ns while the transmission remains above 10\%. The
use of two isotopes allows the delay to be increased or decreased by
optical pumping with a second laser, producing rapid tuning over a
range of more than 40\% of the initial delay at 110$^{\circ}$C. We
investigate the frequency and intensity ranges in which this delay
line can be realised. Our observations are in good agreement with a
numerical model of the system.
\end{abstract}



\maketitle
`Slow light' refers to the propagation of a pulse of light in a
dispersive medium at a group velocity much less than
$c$~\cite{Boyd02}. By its use, optically encoded information can be
controllably delayed without the need for electronic transduction.
This is of great interest for telecommunications, where there is a
need for tunable all-optical delay lines for high-speed optical
signal processing, e.g., buffering optical data packets
~\cite{Ku02}. Additionally, such a system may be included in the
growing repertoire of tools available for quantum information
processing.

To minimise pulse distortion, an optical delay line should have
approximately constant positive dispersion in a spectral region
$\Delta \nu$ of width larger than the inverse of the pulse duration,
i.e., $\Delta \nu > 1/\tau$. The transmission should be high and the
fractional delay (the ratio of the delay $\delta$ to the pulse
duration), which provides a practical metric, should exceed unity,
i.e., $\delta/\tau > 1$.

Narrow spectral features in the refractive index of atomic media due
to light-induced ground-state coherence can result in sub and
superluminal pulse propagation~\cite{Akulshin03} and even `light
storage'~\cite{Phillips01,Liu01}. Using atomic media to produce
optical delay has predominantly exploited the steep dispersion
associated with electromagnetically induced transparency
(EIT)~\cite{Kasapi95, Hau99, Kash99}. While EIT in atomic vapour can
produce extremely low group velocities it has a severe bandwidth
limitation owing to the narrow spectral range over which the
transparency and steep dispersion occurs, making $\delta/\tau
> 0.3$ difficult to obtain. Because of this, it has been suggested
that ultracold atomic samples may be required to achieve large
fractional delays in EIT-based delay lines~\cite{Matsko05}.

In solid-state media, attempts to obtain larger bandwidth include
methods based on spectral hole burning~\cite{Shakhmuratov05} and the
use of gain features such as stimulated Brillouin
scattering~\cite{Okawachi05} and Raman
amplification~\cite{Sharping05} in optical fibres.

An attractive approach to realising a wide-bandwidth delay line
utilises the intrinsic positive dispersion and high transmission
between two absorption lines in an atomic vapour. This has allowed,
Camacho \emph{et~al.} to observe large fractional delays for
light pulses tuned between the $^{85}$Rb hyperfine components of the
$D_2$ line~\cite{Camacho06}. In addition, this technique provides a
high degree of spatial homogeneity in both dispersion and
absorption, allowing the delay of transversely encoded optical
information (images)~\cite{CamachoImage}.

In this paper, we investigate the delay and transmission properties
of optical pulses tuned within the Rb $D_2$ line in a heated vapour
with natural isotopic abundance. Moreover, we modify the dispersion
by optical pumping to either reduce or enhance the number of
interacting atoms on one of the absorption components, allowing
rapid control of the delay.

\begin{figure}[h!b!p!]
\begin{center}
  \includegraphics[width=16cm]{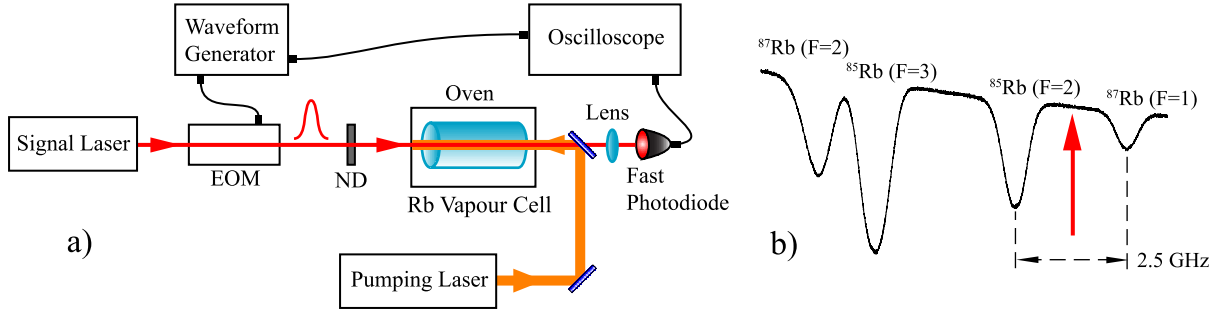}
  \caption{(Colour Online) %
  a) Schematic of the experimental setup. The optical pumping laser is used only for delay tuning. %
  b) Rb $D_2$~line absorption spectrum at room temperature
  where the arrow indicates the region of the signal laser frequency used. %
  }
  \label{Setup}
\end{center}
\end{figure}

The scheme of our experimental setup is shown in
figure~\ref{Setup}a. Signal and optical pumping radiation is
produced using extended cavity diode lasers tuned to the rubidium
$D_2$ (figure~\ref{Setup}b) and $D_1$ lines, respectively. The laser
frequencies are controlled with reference to Doppler-free saturation
spectroscopy performed in auxiliary Rb cells and the spectral
purities are analysed using Fabry-Perot cavities.

Optical pulses of 9.3~ns duration (FWHM) with a repetition rate of
10~MHz are generated from the cw signal laser using an electro-optic
modulator (EOM) triggered by a waveform generator
(figure~\ref{Setup}a). The optical intensity is controlled by a
neutral density filter (ND) before propagation through a 10~cm
vapour cell heated in a thermally insulated oven. The transmitted
pulses are detected using a fast photodiode and recorded on an
oscilloscope.

For rapid tuning of the delay, optical pumping radiation at either
the $^{87}$Rb~(F=1) or (F=2) component of the $D_1$ line is applied,
approximately counter-propagating to the signal beam. A lens
minimises spatial deviation of the signal beam induced by optical
pumping.

\begin{figure}[h!b!p!]
\begin{center}
  \includegraphics[width=15 cm]{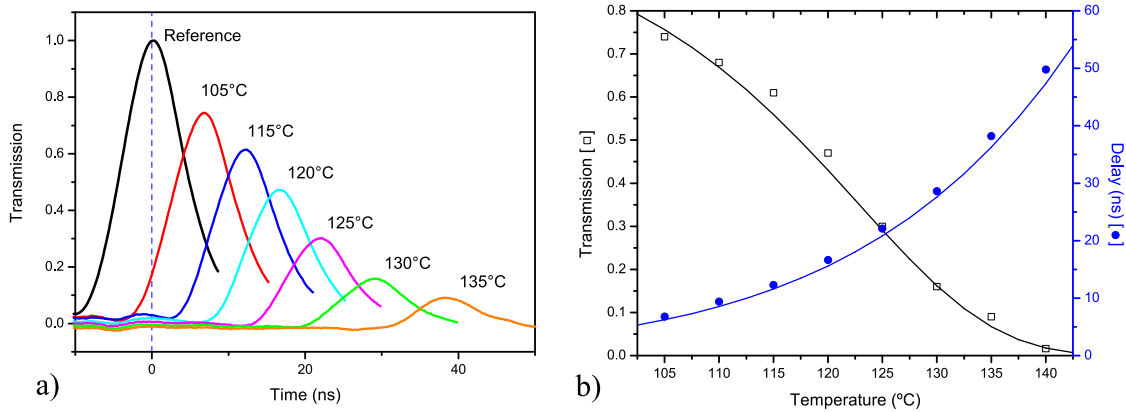}
  \caption{(Colour Online) a) Delayed pulses with increasing temperature. The transmission is
  normalised to a non-interacting reference pulse. %
  b) Observed transmission ($\Box$) and delay ($\bullet$) with numerical predictions (curves).
  }
  \label{PulseDelays}
\end{center}
\end{figure}

The signal laser frequency is tuned to the $D_2$~line at $\lambda =
780$~nm between the $^{85}$Rb~(F=2) and $^{87}$Rb~(F=1) components,
which have a separation of $\sim$2.5~GHz (figure~\ref{Setup}b). The
inherent positive dispersion and low absorption in this broad
spectral region allows large fractional delays with high
transmission. In figure~\ref{PulseDelays}a the observed optical
delays for pulses at the frequency of peak transmission are shown
relative to a non-interacting reference pulse for temperatures
between 105$^{\circ}$C and 135$^{\circ}$C. A fractional delay
$\delta/\tau = 4.3$ was observed for a transmission of 9\% with good
pulse shape preservation, where the pulse duration narrowed by less
than 10\%. It should be noted that the fractional delay is limited
in these experiments by the pulse duration we are able to generate.
The observed delay and transmission are plotted against temperature
in figure~\ref{PulseDelays}b, along with our numerical predictions.

For our numerical predictions, we model the absorption coefficient
$\alpha(\omega)$ and the real part of the refractive index
$n(\omega)$ of the Rb $D_2$~line using a convolution of profiles
arising from homogeneous and inhomogeneous broadening mechanisms.
The homogeneous profile includes contributions from natural
broadening, collisional broadening~\cite{AkulshinCol} and power
broadening and is convolved with the thermal Gaussian Doppler
profile. The group velocity is then approximated using the first
derivative of $n(\omega)$ with respect to $\omega$,
\begin{equation}\label{vg}
v_g = \frac{c}{n(\omega_0) + \omega_0 \frac{\partial
n(\omega)}{\partial \omega}} = \frac{\partial \omega}{\partial k}.
\end{equation}


At the frequency of peak transmission between the two absorption
lines, the probability for resonant interaction via the Doppler
shift is small, even at the temperatures used in these experiments.
Instead, interaction occurs mainly through the broad wings of the
homogeneous component of the profile. For example, at
$T=110^{\circ}$C the probability of an atom belonging to a
velocity class with a detuning of 1~GHz from resonance and an optical bandwidth of 110~MHz
(appropriate for our pulse duration) is about $10^{-4}$. Using an
estimated number density of $10^{13}$ cm$^{-3}$ (based on
Ref.~\cite{SteckRb87} and taking into account the natural isotopic
ratio), the optical depth $\alpha L$ is 0.4 for a 10~cm cell.

\begin{figure}[h!b!p!]
\begin{center}
  \includegraphics[width=12cm]{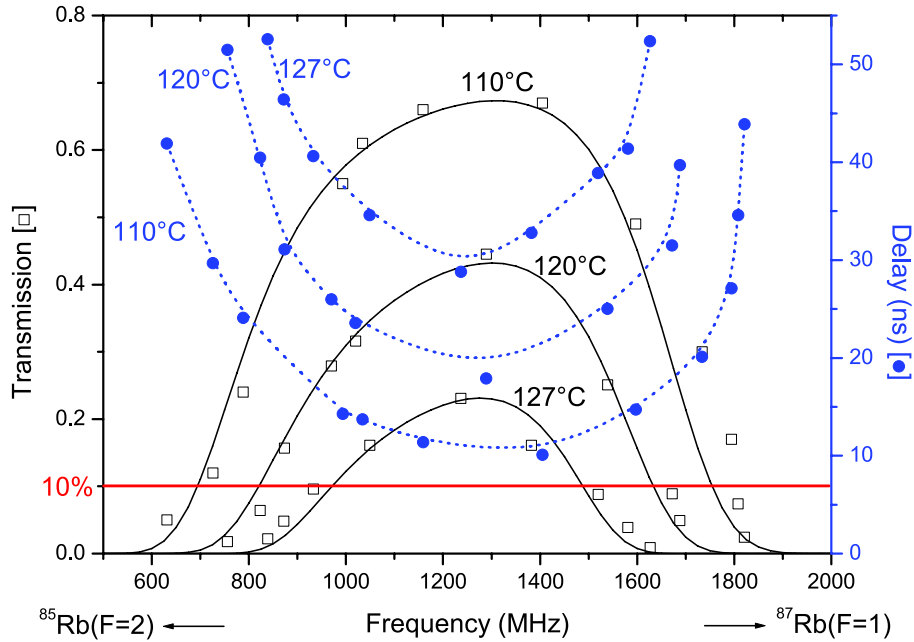}
  \caption{(Colour Online) a) Frequency dependence of the pulse delay
  and transmission for a range of temperatures. The frequency is measured
  from the absorption peak of the $^{85}$Rb~(F=2) component.
  The points are experimental observations and the solid curves are
  our numerical predictions for transmission. The saturation parameter $S=$~20, 8 and 4 for
  $T=$~127, 120 and 110$^{\circ}$C, respectively.
  Interpolations (dashed) are included as a guide to the eye for the delay.
  }
  \label{FreqTuning}
\end{center}
\end{figure}

Our pulse bandwidth of $110$~MHz is less than the width of the
transmission window of approximately 1~GHz which allows the
variation in $v_g(\omega)$ and $\alpha(\omega)$ between the
absorption components to be explored. The frequency dependence of
the pulse delay and transmission at different temperatures is shown
in figure~\ref{FreqTuning}. For 10\% transmission, suggested in
Ref.~\cite{Camacho06} as a practical limit for a delay line, the
usable bandwidth decreases from 1.1~GHz at $110^{\circ}$C to 540~MHz
at $127^{\circ}$C. At the former temperature we expect that a
fractional delay an order of magnitude larger could be achieved with
shorter pulses that utilise the available bandwidth.

The effect of saturation is quantified in our numerical model by the
saturation parameter $S = I/I_{sat}(\omega)$, where
$I_{sat}(\omega)$ is the saturation intensity which is inversely
proportional to the frequency dependent absorption cross section.
This means saturation has little effect on the wings of a
homogeneously broadened line. In contrast, for an inhomogeneously
broadened line, saturation can occur over the entire profile. This
means that although increasing the temperature decreases the usable
bandwidth due to Doppler broadening (figure~\ref{FreqTuning}), at a
given temperature the bandwidth may be increased by increasing the
saturation.

Expression (\ref{vg}) for $v_g$ provides an accurate approximation
for the delay at frequencies near the point of peak transmission.
However, the agreement is found to reduce for frequencies closer to
the absorption peaks. This may be due to higher spectral derivatives
in $n(\omega)$ becoming more significant. It was observed that the
pulse shape suffers little distortion from dispersive and absorptive
mechanisms and remains preserved. This supports the finding in
Ref.~\cite{Camacho06} that these mechanisms compensate one another.


\begin{figure}[h!b!p!]
\begin{center}
  \includegraphics[width=9.5 cm]{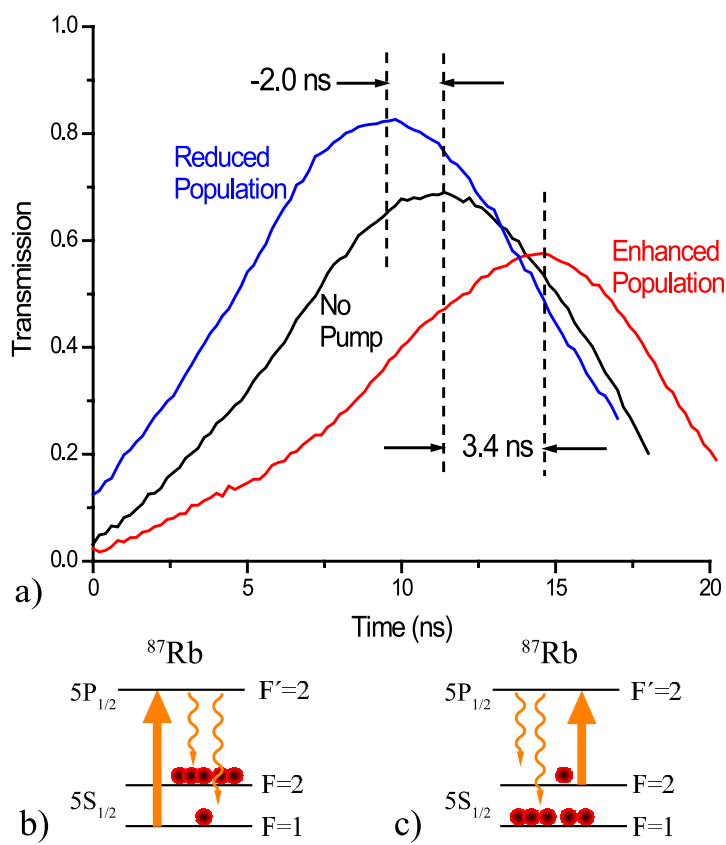}
  \caption{(Colour Online) a) Reduced and increased delay via
  optical pumping, which reduces b) or enhances c) the population of the $^{87}$Rb~(F=1) ground state, respectively. %
  }
  \label{Pumping}
\end{center}
\end{figure}

\begin{figure}[h!b!p!]
\begin{center}
  \includegraphics[width=15cm]{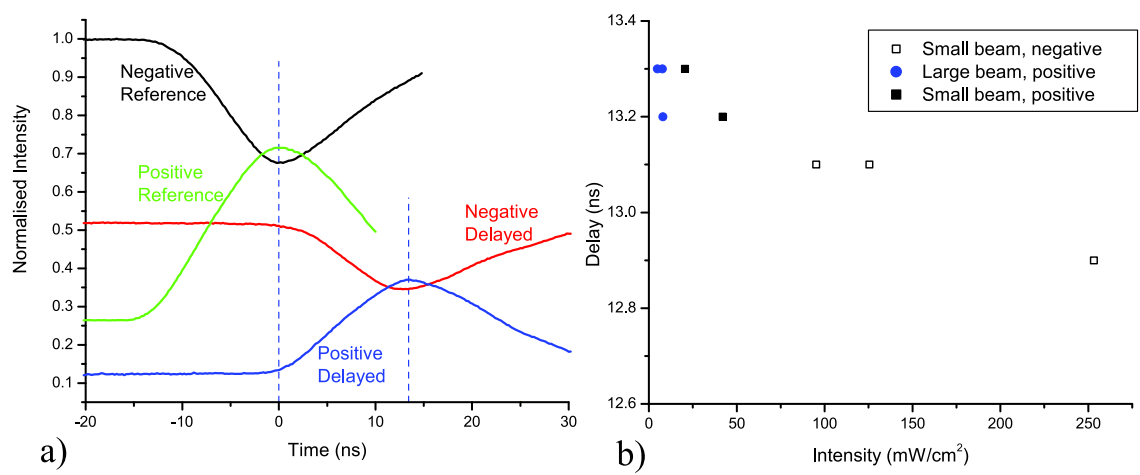}
  \caption{(Colour Online) a) Delay of positive and negative going pulse shapes and b)
observations of delay with average intensity of the pulse train at
$\sim 115^{\circ}$C, where the small and large beam diameters used
were 1 and 3~mm, respectively.
  }
  \label{Intensity}
\end{center}
\end{figure}

Temperature tuning provides a method for changing the delay over a
wide range, but the change is inherently slow as the cell heats or
cools. Fast control of the delay was achieved in
Ref.~\cite{Camacho07} by using two additional laser fields to modify
the dispersion by saturating both absorption lines to reduce the
atomic ground state population. While this approach gives rapid
delay tunability, it produces a relatively small tuning range. In
the present work, where the pulses are tuned between absorption
resonances from different isotopes, hyperfine optical pumping allows
the ground state population of one of the resonances to be strongly
modified with minimal modification of the other. An optical pumping
laser tuned to one of the $^{87}$Rb components of the $D_1$ line is
used to modify the population of $^{87}$Rb~(F=1) ground state atoms
interacting with the signal. Tuning the optical pumping laser to the
$D_1$ $^{87}$Rb~(F=1) or (F=2) component respectively reduces or
increases the population in the F=1 ground state
(figure~\ref{Pumping}b~and~c). Pumping on the $D_1$ line is more
efficient than on the $D_2$ line as it has no cycling transition.
Moreover, the optical depth is less both for this line and for the
$^{87}$Rb component, giving greater longitudinal pumping
homogeneity. In this manner, at $110^{\circ}$C the delay was reduced
by approximately 17.5\% or increased by 25\% of the unmodified delay
(figure~\ref{Pumping}a). The delay reconfiguration time is mainly
limited by atomic time of flight, which is expected to be of the
order of microseconds. A range of delay tuning greater than the
pulse duration should be possible with shorter pulses.

Measurements of the delay of optical pulses with very low intensity
have previously been performed with an average of less than one
photon per pulse~\cite{CamachoImage}. In this work however, we are
interested in establishing high intensity limits, which we do by
using both positive and negative pulse shapes
(figure~\ref{Intensity}a). A negative pulse shape is an interval of
reduced intensity on a relatively large optical DC background.
Negative pulse shapes are found to exhibit similar delay to positive
pulses of comparable intensity, and they may be useful for other
applications in that the signal to noise ratio can be higher.
Furthermore, such pulses may be of interest in experiments involving
atomic or optical coherence. By neutral density filtering and
adjusting the beam waist, the delay was measured for a range of
values of the intensity of the 10~MHz pulse train
(figure~\ref{Intensity}b). It is observed that the delay decreases
with increasing intensity by approximately 1.8~ps/(mW/cm$^2$). We
attribute this to the increase in saturation of the atomic medium,
which has the effect of reducing the dispersion.



In conclusion, optical pulses of 9.3~ns duration (FWHM) with
frequency tuned between the $^{85}$Rb~(F=2) and $^{87}$Rb~(F=1)
components of the $D_2$ line were delayed in a 10~cm vapour cell at
$135^{\circ}$C with low distortion by more than 40~ns (fractional
delay 4.3) and with approximately 10\% transmission. The delay
arises from the intrinsic positive dispersion between the two
absorption peaks. In this experiment the fractional delay was
limited by the pulse duration, but should be ultimately limited by
the $\sim$1~GHz transmission window, making a fractional delay of 40
possible.


The dependence of delay, transmission and usable bandwidth with
temperature and frequency was investigated. With increasing
temperature and atomic density the delay increases and the
transmission reduces. This trend also applies as the optical
frequency is tuned closer to one of the resonances. A reduction in
usable bandwidth was measured with increasing temperature. In
addition, the delay was found to reduce with increasing intensity.
This was observed using negative pulses, which were delayed in a
similar manner to positive pulses.



In contrast to EIT-based delay lines this technique provides the
large bandwidth necessary for delaying short optical pulses and also
operates at both low and high intensity levels.

Using the spectral region between absorption components of different
isotopes for an all-optical delay line allows optical pumping with a
single laser to modify the interacting population of one absorption
component without modifying the other. Rapid tuning of the delay was
obtained in this way over a range more than 40\% of the unmodified
pulse delay at $110^{\circ}$C.

Finally, we note that such broadband delay lines may be used to
delay many forms of optical quantum information encoding such as
weak coherent pulses or squeezed states.



\end{document}